\documentclass[12pt,preprint]{aastex}

\shorttitle{Mid-Infrared Observations of NGC 3576}
\shortauthors{Barbosa et al.}

\begin{document}


\title{Gemini Mid-Infrared Imaging of Massive Young Stellar Objects in NGC 
3576\footnote{Based on observations obtained at the Gemini
Observatory, which is operated by the Association of Universities for
Research in Astronomy, Inc., under a cooperative agreement with the
NSF on behalf of the Gemini partnership: the National Science
Foundation (United States), the Particle Physics and Astronomy
Research Council (United Kingdom), the National Research Council
(Canada), CONICYT (Chile), the Australian Research Council
(Australia), CNPq (Brazil) and CONICET (Argentina)} }


\author{C. L. Barbosa \& A. Damineli}
\affil{Instituto de Astronomia, Geof\'{\i}sica e Ci\^encias Atmosf\'ericas, Universidade de
S\~ao Paulo; R. do Mat\~ao 1226, 05508-900, S\~ao Paulo, Brazil}
\email{cassio@astro.iag.usp.br}
\email{damineli@astro.iag.usp.br}

\author{R. D. Blum}
\affil{Cerro Tololo Interamerican Observatory, Casilla 603, La Serena, Chile}
\email{rblum@ctio.noao.edu}

\and

\author{P. S. Conti}
\affil{JILA, University of Colorado Campus Box 440, Boulder, CO, 80309-0440}
\email{pconti@jila.colorado.edu}


\begin{abstract}
We present a mid-infrared study of NGC 3576. The high-resolution
images were taken at the Gemini South Observatory through narrow and
broad band filters centered between 7.9 $\micron$ and 18
$\micron$. The nearly diffraction limited images show IRS 1 resolved
into 4 sources for the first time in the 10 $\micron$ band. The positions of
the sources are coincident with massive young stellar objects detected
previously in the near infrared. The properties of each object, such
as spectral energy distribution, silicate absorption feature, color
temperature and luminosities were obtained and are discussed. We also
report observations of two other YSO candidates and the detection of a
new diffuse MIR source without a NIR counterpart. We conclude that
none of these sources contributes significantly to the ionization of
the \ion{H}{2} region. A possible location for the ionization source
of NGC 3576 is suggested based on both radio and infrared data.
\end{abstract}




\section{Introduction}

Massive stars are responsible for many important phenomena in
galaxies. They alter their environment through the emission of high
energy radiation and the deposition of momentum and mechanical energy
into the interstellar medium through powerful winds. At the end of
their lives, massive stars explode as supernova enriching the
interstellar medium and causing shocks that may trigger the formation
of new stars. The life and death of massive stars have a profound
impact on both local and galactic scales.

Massive stars are believed to form in warm dense cores of molecular
clouds. The earliest stage of a star in the process of forming is
known as `Prestellar Core' (PSC) \citep{church02}. PSCs have not yet
formed a central protostar and will not be detected in the near
infrared (NIR) or at radio wavelengths. Moreover PSCs have
temperatures of only $\sim$20 K and their spectral energy
distributions (SEDs) peak in the far infrared (FIR) at $\sim$200
$\micron$ \citep{garay99}. The next phase of evolution toward the
main sequence is known as the Hot Core (HC) phase. In this stage, a
compact, dense and warm molecular cloud core is believed to harbor a
massive protostar in a process of rapid accretion, probably surrounded
by an equatorial accretion disk \citep{kurtz00}. HCs are not detected
at wavelengths shorter than $\sim$10 $\micron$, their SEDs are broader
than a single temperature black body distribution, but they have a
pronounced peak at $\sim$100 $\micron$. Since the protostar
experiences a process of rapid massive accretion it cannot produce a
detectable \ion{H}{2} region \citep{osorio99}.  After the accretion is
over or greatly diminished, the Lyman Continuum photons emitted by the
massive star ionize the surrounding gas and a ultra-compact \ion{H}{2}
(UC\ion{H}{2}) region is formed. UC\ion{H}{2} regions have SEDs also
peaking at $\sim$ 100 $\micron$, but they now can be detected at
shorter wavelengths, such as 1-2 $\micron$ \citep{church89b,hanson00}. The
ionized gas is surrounded by a warm dust cocoon which makes
UC\ion{H}{2} regions bright sources in the mid infrared (MIR), as
well. As the star evolves, the UC\ion{H}{2} region expands, the natal
gas and dust are swept away by intense stellar winds. Eventually, the
ionizing star becomes visible, but it may have moved 10-15\% along its
evolutionary track (i.e after the zero age main sequence) at this
stage \citep{garmany94}, and important questions regarding the
formation of massive stars and their environment cannot be directly
addressed anymore.

NGC 3576 harbors at least a dozen intriguing objects with color
indexes $H-K$$>$2, identified by \citet[hereafter FBDC]{lys02}. The
$K$-band spectra of some of the brighter sources do not show any
photospheric features. Moreover, the CO 2.3 $\micron$ bandhead is seen
in emission or absorption in the spectra of 4 of these
objects. Although the presence of this feature in emission or
absorption has been explained by a variety of mechanisms, such as
circumstellar disks, stellar or disk winds, magnetic accretion,
instabilities in the inner regions of accretion disks of low mass
stars or free-falling gas along field lines (see the references in
FBDC), disk emission is the most preferred model, at
least for low mass stars. The case is not yet clear for massive stars.
The CO bandhead was also found in emission in massive stars in M17 by
\citet{hanson97} in low resolution spectra. Optical spectra of some of
these same stars suggest they do have circumstellar disks. 

The ``$H-K$ excess" objects in NGC~3576 may be even younger than the M17
objects, which are visible in the $I$ band \citep{hanson97}
and are believed to be massive young stellar objects (YSOs)
perhaps surrounded by a thick accretion disk. There are analogous
massive objects with strong NIR excess in W31 \citep{blum01}, W42
\citep{blum00}, and W49 \citep{cont01}.

Four sources - \#48, \#50, \#60 and \#60b (all sources with ``\#" in the
present paper follow the FBDC nomenclature) - were found at the position
of the MIR source IRS 1, identified by \citet{frogel74}. The $K$-band
spectrum of \#48 does not show any photospheric lines and moreover,
source \#50 was not detected at $J$ and $H$ bands. For this
reason no further information, such as spectral type could be derived
from NIR data. However, at this evolutionary stage, crucial stellar
parameters, like luminosity and hence the stellar mass, can be
inferred by measuring their fluxes in the MIR.  This spectral regime
is also suitable to study the environment where the YSOs are
forming. The spatial distribution of dust can put constraints on the
geometry of accretion since accretion via a disk needs large amounts of
gas (which is mixed to the dust) concentrated in a small region.

NGC 3576 (also known as G291.3-0.71 or RCW57) is a giant \ion{H}{2}
region located in the Galactic plane at a kinematic distance 2.8 ($\pm
0.3$) kpc (FBDC). It was observed at radio wavelengths by
\citet{mcgee68}, \citet{goss70}, \citet{wilson70}, \citet{mcgee81} and
\citet{depree99}. Methanol and water masers were detected by
\citet{caswell95} and \citet{caswell89}, respectively. NIR photometry
was performed by \citet{moor81}, \citet{moneti92}, \citet{persi94}
and FBDC who presented a deep NIR study of NGC 3576.  \citet{frogel74}
originally observed the region in the MIR. Five sources were identified
and the brightest, IRS 1, was unresolved with a 7$\arcsec$ diaphragm.
Later MIR observations of NGC 3576, from \citet{moor81}, \citet{lacy82},
\citet{persi87} and more recently, \citet{walsh01}, were also unable
to resolve the source IRS 1.

In this paper, we present high-resolution MIR observations of 3 selected
fields in NGC 3576, taken through the atmospheric windows near 10 $\micron$
and 18 $\micron$, in $\S$ 2. The data and the discussion of the results are
presented in the $\S$ 3. The conclusions are summarized in $\S$ 4.

\section{Observations and Data Reduction}

The data were obtained at the Gemini South Observatory 8-m
telescope on 2001 November
30th and December 4th and 6th in service mode with the University of
Florida OSCIR\footnote{This paper is based on observations obtained
with the mid-infrared camera OSCIR, developed by the University of
Florida with support from the National Aeronautics and Space
Administration, and operated jointly by Gemini and the University of
Florida Infrared Astrophysics Group.} MIR camera. The camera employs
a Rockwell 128$\times$128 pixel SI:As BIB detector, the plate scale
at Gemini was 0.0859\arcsec /pixel; the total field of view (FOV) of
the array was 11\arcsec$\times$11\arcsec.

Flux calibration was performed using the MIR standard star $\alpha$
CMa observed during the night as part of the baseline calibration
program. Air mass differences between the standard star observations
and the target observations were $<$0.3 resulting in an uncertainty of
approximately 10\%. Table \ref{stdlog} presents a summary of the observations, as
well as the filter parameters. Sky and background subtraction were
achieved by the standard chop-and-nod technique. Sky images were taken
$\sim 15 \arcsec$ to the north of the selected fields. Images were
processed using the OSCIR reduction package running under
IRAF\footnote{IRAF is distributed by the National Optical Astronomy
Observatories.}  environment, including the flat field correction. The
processed images were also corrected for bad pixels and finally were
flux calibrated. The photometry was performed assuming a Gaussian PSF
model fitted to the objects.

Images were obtained through the $N$-broad band at 10.5 $\micron$ and
the narrow-band filters at 7.9, 9.8, 12.5 and 18.2 $\micron$. The 40 seconds
exposure time was the minimum value to allow a complete chop-nod cycle. Based on
pre-commissioning sensitivity, this value would result in a narrow-band S/N of
$\sim$3 for a 100 mJy point source.

The detection limit (S/N$\sim$1) is $\sim$100 mJy in all bands, except at
18.2 $\micron$, which is $\sim$900 mJy. Three fields were observed: the
region associated with IRS 1 (in all bands), the region centered on
source \#95 (at 7.9, 9.8, 12.5 and 18.2 $\micron$ bands) and field
where sources \#52 and \#54 are located (only at the 12.5 $\micron$
band).

\section{Results and Discussion}

The first field observed was centered at the position of IRS 1, where
3 NIR sources were found by \citet{persi94} and FBDC: \#48, \#50 and
\#60. The second field observed is located $\sim 10\arcsec$ east of
IRS 1. This field hosts 2 YSO candidates (\#95 and \#83) and a
late-O/early-B star (\#85) identified in the color-magnitude diagram
presented by FBDC (their Figure 3). This second set of images,
however, has lower S/N ratio and for this reason we could measure the
flux only for source \#95 in the 7.9, 9.8 and 12.5 $\micron$
bands. The third field was imaged only in the 12.5 $\micron$ band and
for this reason will not be presented. It is located $\sim 25\arcsec$
southern of IRS 1 and hosts one YSO candidate (\#52) and an object
detected by FBDC only in the $K$-band (\#54). A third object was
detected in this field as a negative pattern, its position relative to
the extend emission seen at the top of image, led us identify this
object as source \#73 caught in the sky beam of chopping procedure, 15
$\arcsec$ north.

\subsection{The Images}
\subsubsection{Sources \#48, \#50 and \#60}

\begin{figure}
\epsscale{0.7}
\plottwo{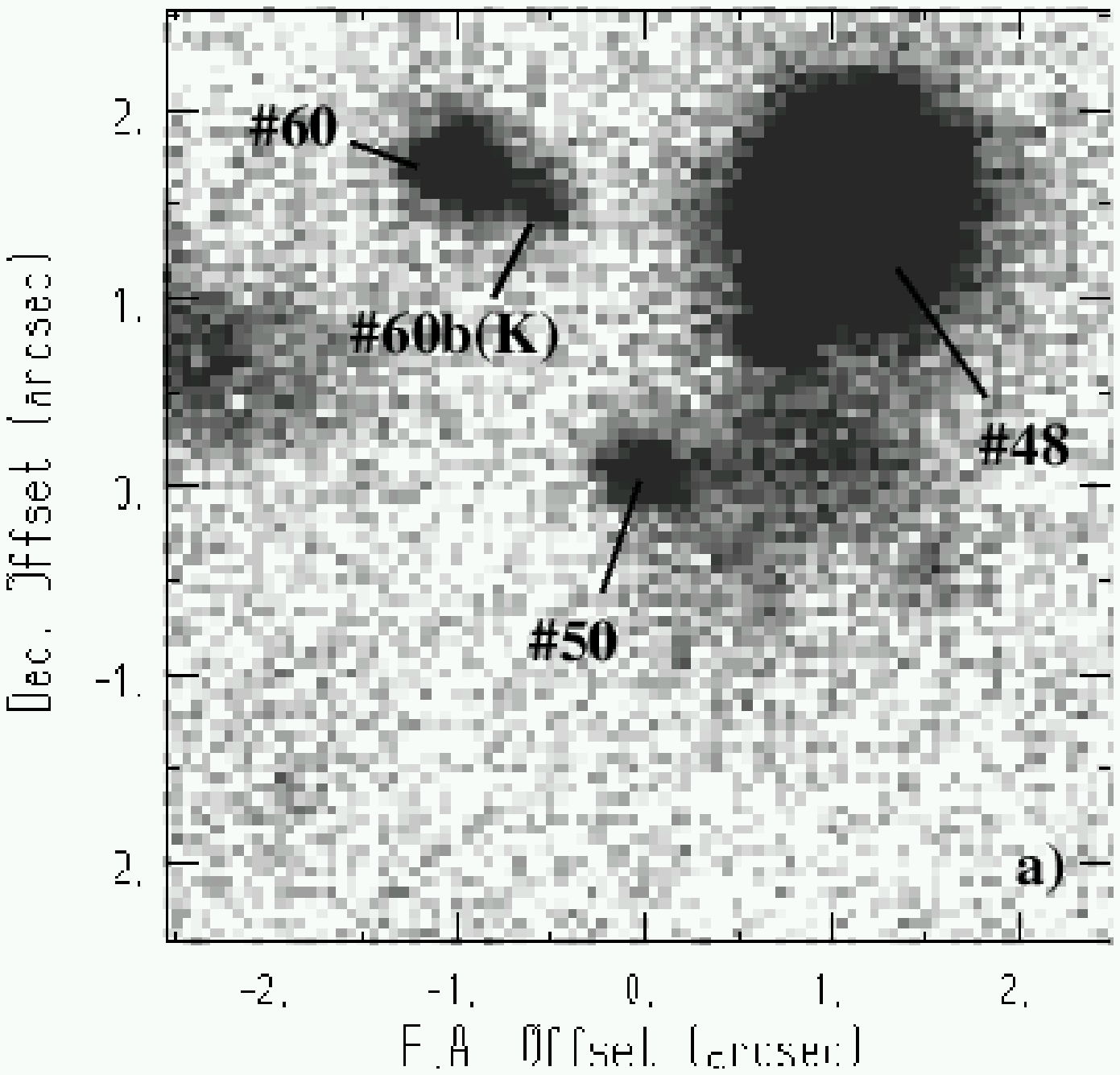}{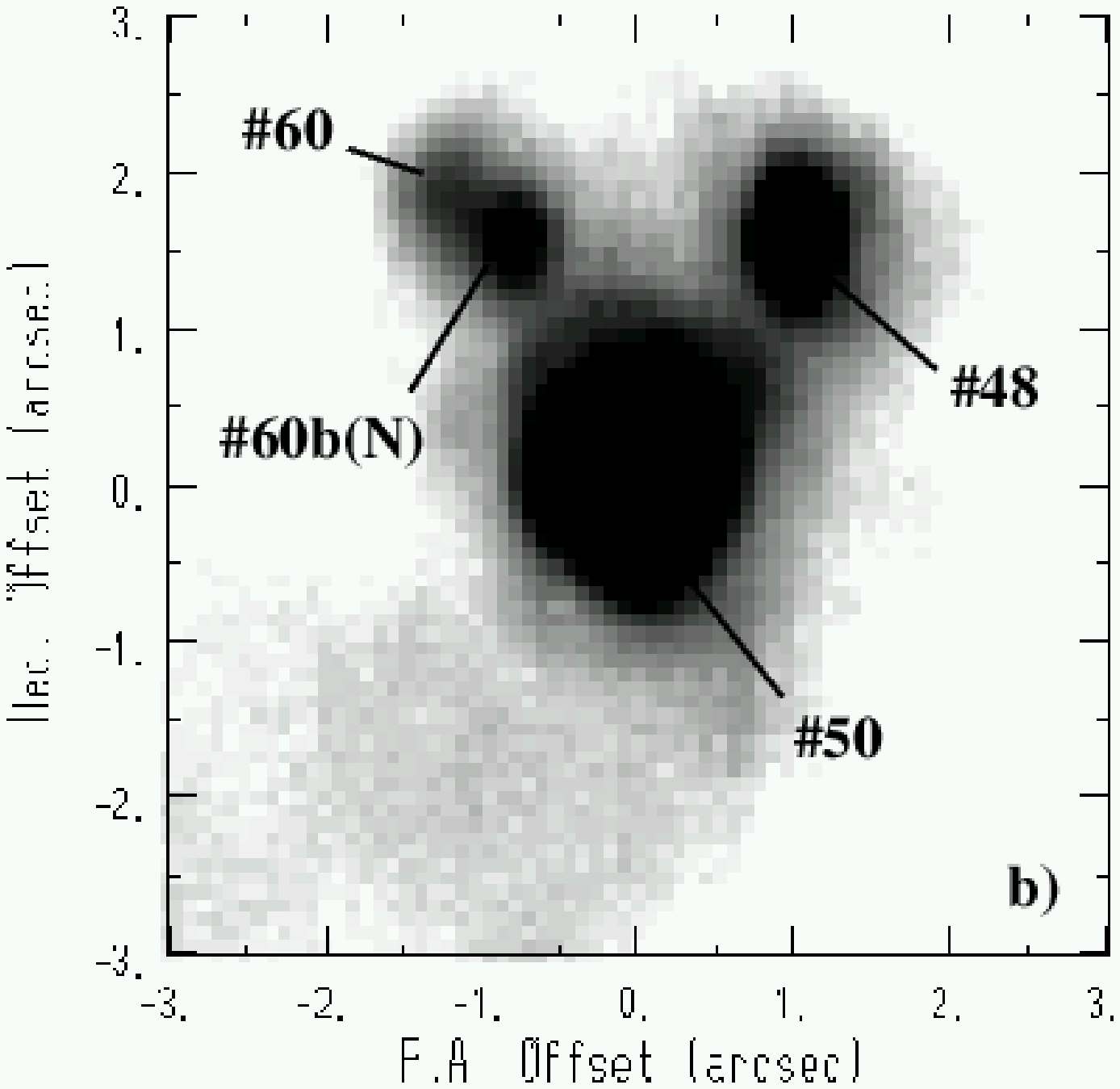}
\caption{a) $K$-band image of IRS 1 taken with the PHOENIX Acquisition
Camera \citep{blum03}. b) OSCIR $N$-band image of IRS 1. The source numbers
are from FBDC; 1$\arcsec$ corresponds to 0.014 pc.
\label{figN}}
\end{figure}

Sources in IRS 1 are presented in Figure \ref{figN}. Panel (a) shows
the $K$-band image of sources \#48, \#50 and \#60 taken with the PHOENIX
acquisition camera under good seeing ($<$0.3\arcsec), the average PSF is
$\sim0.35\arcsec$ and the camera plate scale is 
0.055$\arcsec$/pixel.\footnote{www.gemini.edu/sciops/instruments/phoenix/phoenixindex.html}
The sources are labeled according to FBDC. Figure \ref{figN} (b) displays the
$N$-band image of the same region showing the sources in IRS 1 resolved
for the first time at 10 $\micron$ and identified with the NIR
counterparts (again, adopting the source numbers of FBDC). The
brightest MIR source is \#50, followed by \#48, opposite from the
situation seen in the $K$-band image, which shows source \#48 brighter
than source \#50. For this reason, source \#48 was associated with IRS
1 by previous authors. Both images also show the double nature of
source \#60. The companion of source \#60, hereafter \#60b, is located
$\sim0.4\arcsec$ away to southwest. Like source \#50, it is brighter
at longer wavelengths.

\begin{figure}
\epsscale{0.5}
\plotone{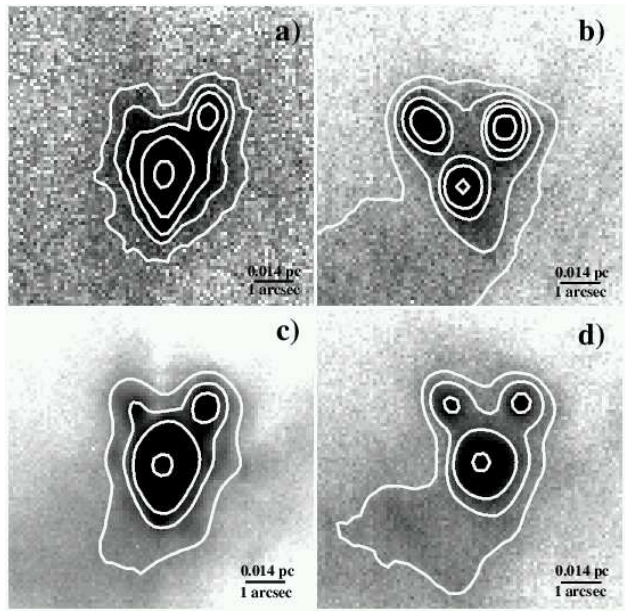}
\caption{a) Contoured image of IRS 1 for the 7.9 $\micron$ filter. The
contour levels are at 2.5, 5, 10, 30 and 300 mJy. b) Contoured
image of IRS 1 for the 9.8 $\micron$ filter.  The contour levels are
at 3, 4.5, 7.5, 10 and 23 mJy. c) Contoured image of IRS 1 for the 12.5
$\micron$ filter. The contour levels are at 10, 20, 40 and 500 mJy.
d) Contoured image of IRS 1 for the IHW18 (18.2 $\micron$) band. Contour
levels are at 40, 65, 100 and 350 mJy.\label{mos-irs1}}
\end{figure}

The images taken through the MIR narrow-band filters are presented in
the Figure \ref{mos-irs1}.  Every image shows persistent artifacts
located at the upper and lower right corners produced by the chop and
nodding procedure. A careful inspection of the sky beam images did not
show any point source object, however the artifacts could be produced
by weak diffuse sources in the chopped sky beam. A logarithmic gray
scale was used to emphasize the low level extended emission where the
sources are embedded. The elongated shape in the N-S direction seen in
the images at 7.9 $\micron$ and 12.5 $\micron$ (Figure \ref{mos-irs1}a
and \ref{mos-irs1}c, respectively) was produced by non optimum tuning
of the primary mirror active optics system.  A careful inspection of the
images, especially the image taken at 9.8 $\micron$ (Figure
\ref{mos-irs1}b), shows \#60 elongated in the NE-SW direction. In this
case, it is the effect of its near companion, as seen in the $K$-band
image. It can be noted also, that source \#60 is undetected in 7.9 and
12.5 $\micron$, but it is clearly seen in the 9.8 $\micron$ image.
This brightening at 9.8 $\micron$ comes from the emission of dust
silicate grains in low density regions surrounding source \#60.

The case of source \#60b(N) and \#60b(K) is somewhat
different. Comparing the images taken at $K$ and $N$, we note that its
centroid, relative to the centroid of other sources in the field, is
shifted by $\sim0.2\arcsec$. Both images were supposed to be taken under the
same orientation, but a slight rotation could lead us to a mistake in the
identification of the sources. The positions of the sources, relative to source
\#50, are listed in the Table \ref{shifts}. They were obtained from the
astrometry of sources in the Figure \ref{figN}. The positions of all sources are
the same (whithin the errors), except the position of source \#60b. This fact
means that the $K$-band image has the same orientation of the $N$-band image.
Therefore we believe sources \#60b(K) and \#60b(N) to be associated.
This shifting in the positions represents $\sim10^{16}$ cm at 2.8 kpc. This
length scale is nearly the same expected for UC\ion{H}{2} regions which have
radii up to $10^{16}$ cm and dust cocoons that are 10 times larger
\citep{church89b}. We speculate that the positional shift is caused by
emission and/or reprocessing of radiation arising from different places in
the source \#60 dust cocoon. $K$-band emission could be radiated by superheated
dust grains located in a dense, self-protective accretion disk, near the star
\citep{church02}. On the other hand, radiation detected at longer
wavelength, such as the $N$ band, is emitted in the outer layers of the
cocoon by warmer dust grains. The spatial resolution of both images,
$\sim0.4\arcsec$ (or $\sim1.6\times10^{16}$ cm) is just high enough to
make this effect evident.

\subsubsection{Sources IRS1:SE and \#95}

\begin{figure}
\epsscale{0.6}
\plotone{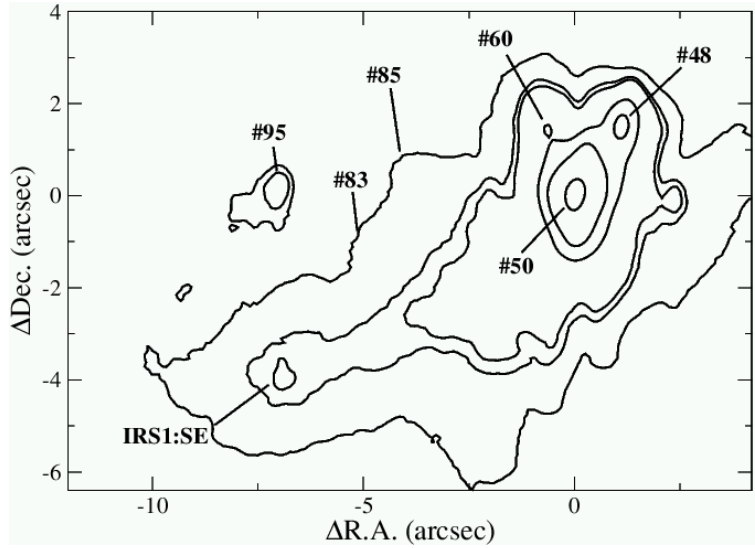}
\caption{Contoured mosaic of IRS 1 and source \#95 taken in the 12.5 $\micron$
band. The contour flux levels are in arbitrary units. A logarithmic scale was
used to emphasize the extended emission around the sources.\label{comp}}
\end{figure}

\begin{figure}
\epsscale{0.7}
\plotone{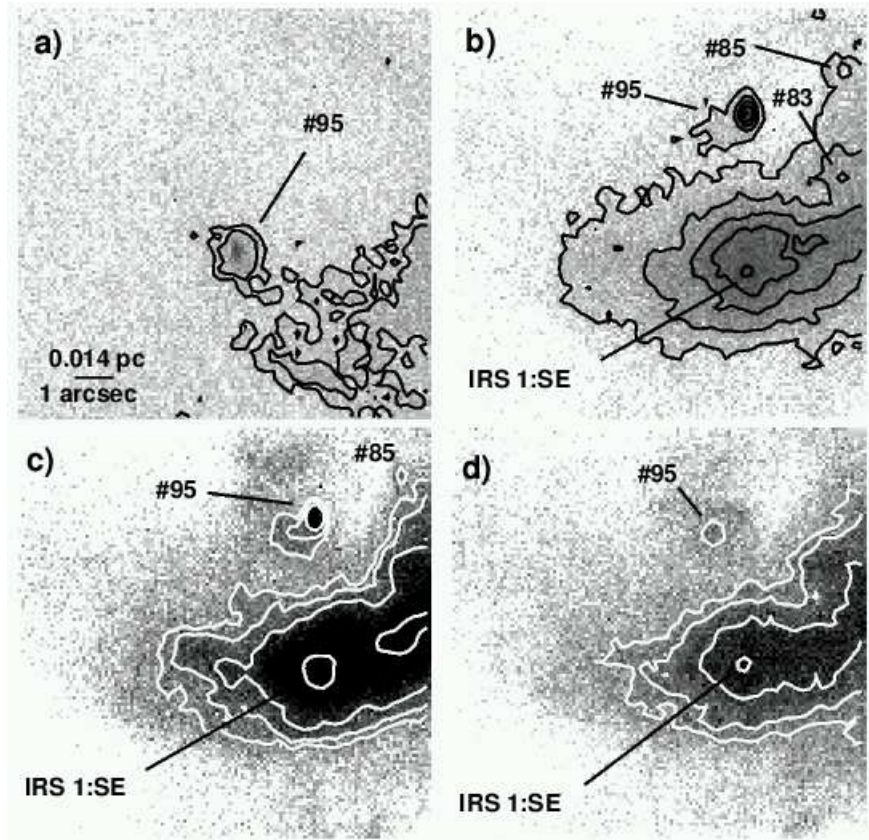}
\caption{a) 7.9 $\micron$ image of source \#95; contour levels
are at 80 and 100 mJy. b) 9.8 $\micron$ image of source \#95. The new
MIR source IRS 1:SE is indicated; contour levels are at 2, 3, 4, 5
and 6 mJy. c) 12.5 $\micron$ image of source \#95; contour levels are at 3.5,
4, 5 and 8.7 mJy. d) 18.2 $\micron$ image of source \#95; contour levels
are at 25, 30, 40, 50 and 60 mJy. \label{mos-95}}
\end{figure}

Source \#95 is presented in Figure \ref{comp}, a contoured mosaic in
the 12.5 $\micron$ band built from the images taken at the positions
of IRS 1 and source \#95. The level curves are in arbitrary units but a
logarithmic scale was used to enhance the extended emission surrounding the
sources. This figure is composed of 2 adjacent
images with a small overlapping area. The positions of sources \#83
and \#85 are indicated for reference, as well as the position of a
previously unknown MIR source, detected 8$\arcsec$ southeast of IRS
1. This source does not have any NIR counterpart to be correlated, so
we named it IRS 1:SE. Images of source \#95 and IRS 1:SE are shown in
the Figure \ref{mos-95}, the extended emission seen around the sources comes
from the warm dust distributed in the intra-cluster medium. Source \#50 has a
PSF FWHM which is statistically indistinguishable from that of the unresolved
standard star. IRS 1:SE is an extended source but it has a
prominent peak as intense as source \#95. For this reason and the fact that
IRS 1:SE is located in low density end of the ``tail" of the extended
emission of \#50 we believe it is an embedded source instead of a
bright knot in the cloud. The position of the peak emission is $\alpha$= 
11h11m54.8s and $\delta$= -61$\arcdeg$18$\arcmin$26$\arcsec$ (J2000). Source
\#95, instead, closely resembles a cometary UC\ion{H}{2} in the
classification scheme of \citet{church89b}.

\subsection{Photometry}

\begin{figure}
\epsscale{0.7}
\plotone{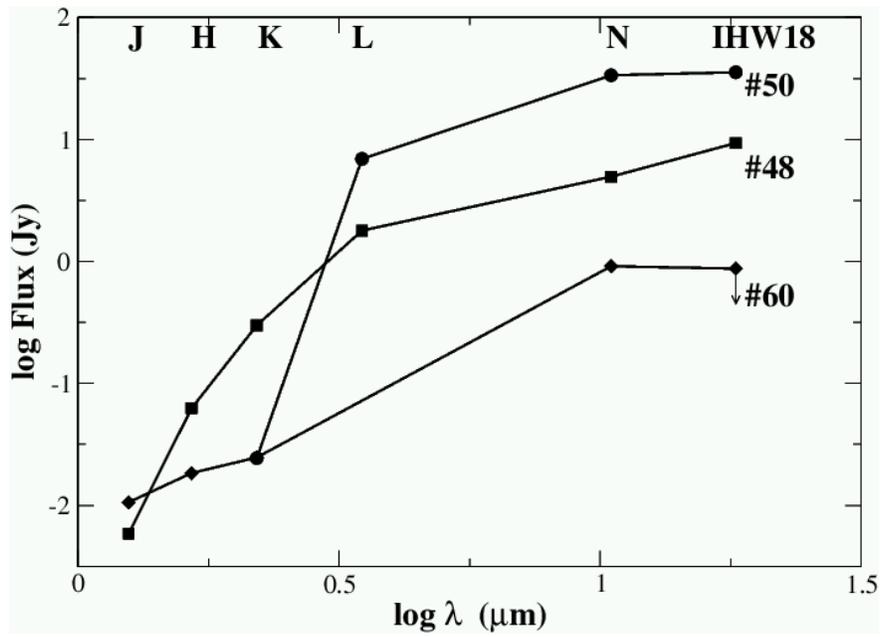}
\caption{NIR to MIR broad--band spectral energy distribution of IRS
1. Squares correspond to source \#48, circles correspond to source
\#50 and diamonds to source \#60. $JHK$ fluxes are from FBDC, L flux
from \citet{moneti92} and MIR fluxes from this work. Error bars are
smaller than symbols.\label{nirsed}}
\end{figure}

Photometry was performed by fitting a Gaussian PSF to the objects
detected in the images, except IRS 1:SE. In this case the fluxes were
obtained through a circular aperture of 1.5$\arcsec$ radius centered
at the peak of IRS 1:SE emission.  The $N$-band flux extracted from the
calibrated image is given in the Table \ref{nir-mir} along NIR
fluxes. The $JHK$ fluxes were calculated from the magnitudes reported by
\citet{lys01} and the $L$ flux is from \citet{moneti92}. The magnitudes were
converted into fluxes adopting the zero points from
\citet{bessell88}. SEDs in the range 1.25 to 18.2 $\micron$ are
plotted in the Figure \ref{nirsed}.  The SEDs show a flat spectrum for
source \#48 and a spectrum rising toward longer wavelengths for source
\#50. The 18.2 $\micron$ flux for sources \#60 and \#60b might be the
fluxes of sources \#60 and \#60b together, since at this wavelength
the sources remain unresolved.

\begin{figure}
\epsscale{0.65}
\plotone{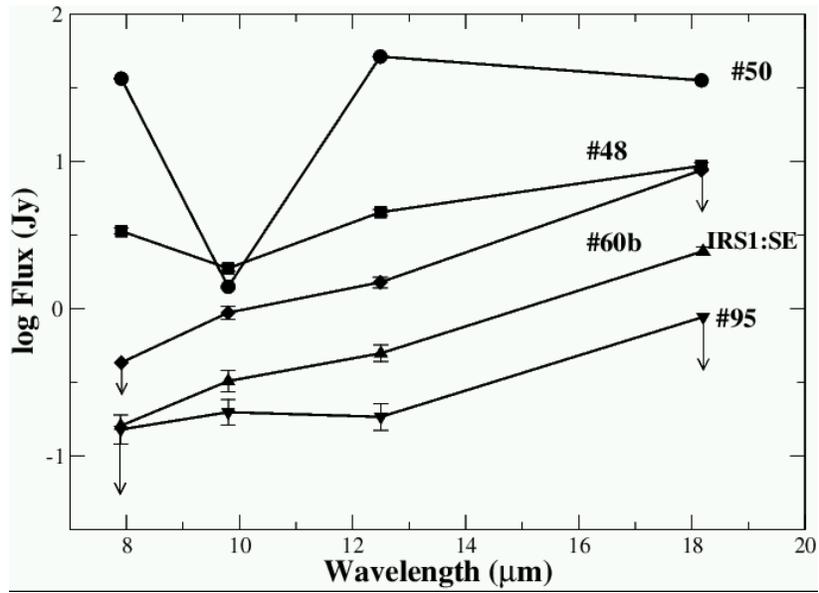}
\caption{Narrow--band MIR spectral energy distribution of sources in
NGC 3576. Squares correspond to source \#48, circles correspond to
source \#50 and diamonds to source \#60. Upward facing triangles
correspond to source IRS 1:SE (Jy arcsec$^{-2}$ for this extended
source), and downward triangles correspond to source \#95. The arrow at 7.9
$\micron$ is the upper limit of IRS 1:SE and it is connected to the up
triangle. The arrow at 18.2 $\micron$ is the upper limit of source \#60b.
\label{mir-sed}}
\end{figure}

Fluxes measured for each source at MIR wavelengths are presented in
Table \ref{mir}. The upper limits reported in this table (except for the
sources \#60 and \#60b) represent the fluxes of the local background emission.
They were obtained by integrating the flux of the extended emission at the
position of the source over a circular aperture of 0.6$\arcsec$. Therefore they
represent the minimum flux that the source indicated should have to be detected.
The SEDs corresponding to the fluxes obtained with the
narrow-band filters are shown in the Figure \ref{mir-sed}. The
expected level of free-free MIR emission was obtained by
extrapolating the 3.4 cm flux from \citet{depree99}, assuming
$S_{\nu}\propto\nu ^{-0.1}$, after deconvolving the
$7\arcsec\times7\arcsec$ radio beamwidth. The free-free emission represents
0.2\% of the $N$-band flux of source \#48 and \#60b, 0.02\% for \#50, 0.05\%
for \#95 and 0.1\% for IRS 1:SE.

While there is no reason to expect the standard classification of low mass YSOs
to be similar to that for high mass YSOs, we can calculate similar spectral
indices for the present sources to compare with those of lower mass stars.
The 2.2-12.5 $\micron$ spectral indices
$\alpha=\frac{d\log{\lambda F_{\lambda}}}{d\log{\lambda}}$ \citep{lada87} for
sources \#48, \#50 and \#95 which we have measured the fluxes at 2.2 and 12.5
$\micron$ are reported in the Table \ref{spt}. Sources \#48 and \#50 have
indices like low mass Class I object ($\alpha_{\#48} = 0.6; \alpha_{50} = 3.4$)
and source \#95 has a similar index as low mass Class II object 
($\alpha_{\#95} = -0.8$), according to the classification scheme of
\citet{greene94}.

\begin{figure}
\epsscale{0.8}
\plotone{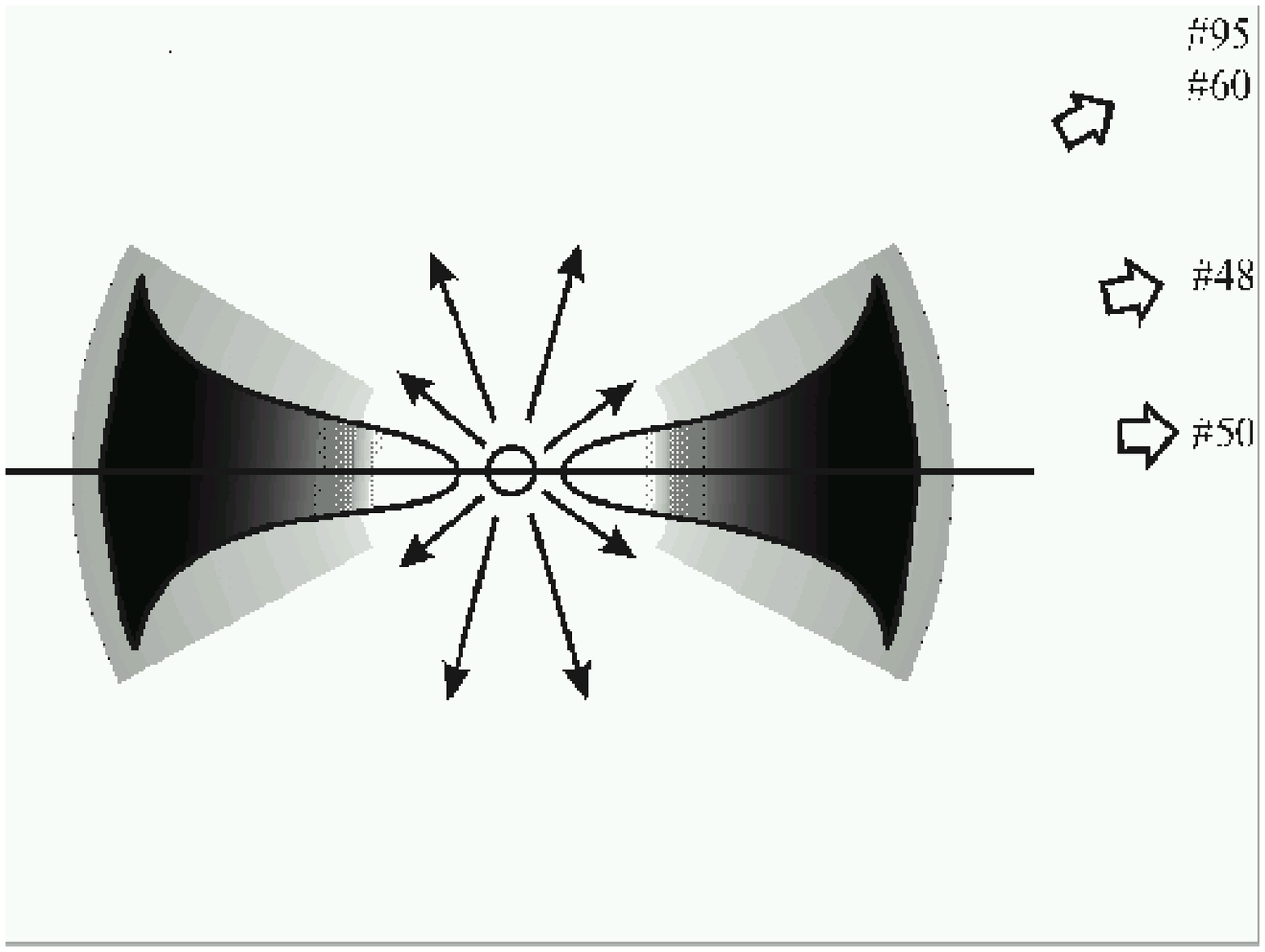}
\caption{Schematic of the geometry of birth material surrounding the
detected sources. Viewing such a source from different lines of site
could produce the different observed SEDs for the NGC3576 objects.
The torus is flared due to the action of strong stellar winds
\citep{holl94}. The inner radius of a typical dust shell is
$\sim10^{16}$ cm and the outer radius is $\sim10^{18}$ cm
\citep{church90,faison98}. Superheated dust grains can survive in the
self protective inner regions of the torus at radii smaller than $10^{16}$
cm. The drawing is not to scale; see text for details.\label{torus}}
\end{figure}

The geometry of birth sites of these stars can explain the differences
in the observed SEDs. It might be that the stars are surrounded by
their birth material in the form of a torus seen at different lines of
sight. The radiation escaping along the rotation axis would produce a
dust evacuated region, disrupting the spherical symmetric dust cocoons
making them highly non uniform in density. Therefore, the stellar
luminosities derived from the fluxes reprocessed by the dust can be taken
as lower limits, only. Figure \ref{torus} is a sketch of this scenario.
The light emitted by sources viewed edge-on or nearly edge-on is absorbed
by the torus, consistent with deep dust absorption observed at 9.8 $\micron$
and little or no emission at wavelengths shorter than 10 $\micron$.
However, the emission of superheated dust grains in the inner radius of the
torus (seen as white) could be detected in the $K$ band if it is not
seen exactly edge-on. This would be the case of source \#50. From
above (or below) the torus equatorial plane, the light emitted by the
star would travel through regions with lower density, implying
lower absorption by the dust. From this viewpoint, the light emitted
at NIR wavelengths ($\leq 2.2 \micron$) could be detected, but
photospheric lines would appear veiled by the inner torus
emission. The CO bandhead at 2.3 $\micron$ is detected in emission if
the angle is small (the case of source \#48), otherwise it would be
detected in absorption (like we see in the spectra of the YSO
candidates \#4, \#160 and \#184, for which we do not have any MIR data
yet). Increasing the angle over (or below) the equatorial plane, one
would observe the radiation that crosses regions with even lower
densities (the gray volume around the torus), in this case the dust
absorption feature is absent and the star becomes brighter at wavelengths
shorter than 2 $\micron$. This would be the scenario for sources \#95 and
\#60. Even at this vantage point, the photospheric lines would still
appear veiled by the inner torus emission and/or by radiation reflected
by the lower density dust near the star.

\subsection{Color Temperature, Luminosity and Silicate Absorption}

Observations at two different wavelengths can be combined to determine
the dust color temperature and optical depths \citep{ball96,buizer00}.
Since temperature determines the ratio of blackbody flux densities at any two
wavelengths, color temperature maps can be obtained by simply ratioing two
calibrated images at different wavelengths.

\begin{figure}
\epsscale{0.8}
\plottwo{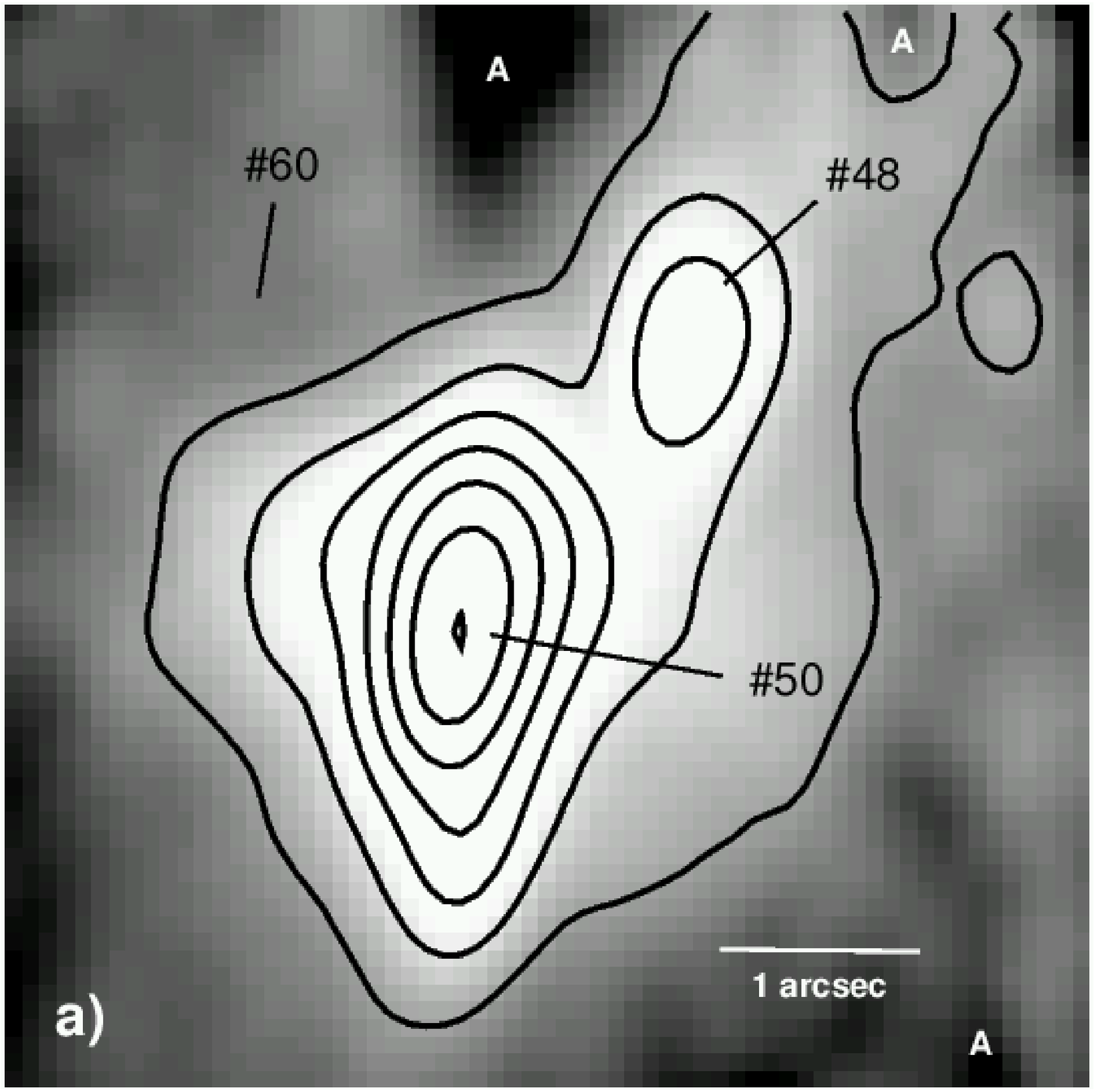}{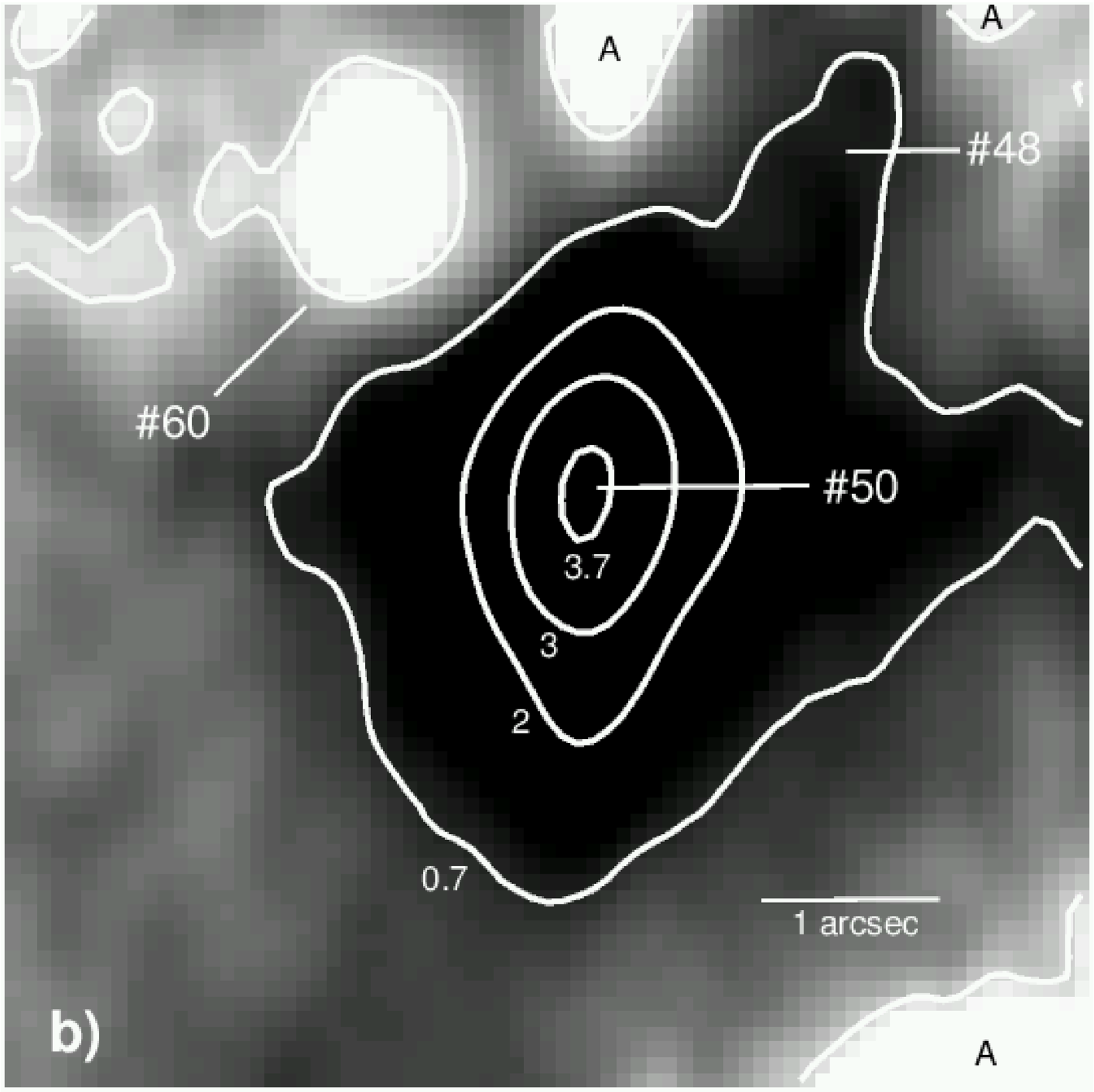}
\caption{a) Color temperature map made from the ratio of the 7.9 and
18 $\mu$m images. The contours are at 160, 180, ..., 280 K. b) Map of the
optical depth of the silicate absorption feature at 9.8 $\mu$m. The darker
regions exhibit the strongest absorption. The rounded contour at the
position of source \#60 marks where the absorption turns into
emission. In both images: ``A" indicates {\it artifact} produced
by the low S/N ratio at this position after ratioing the images.\
\label{ctemp}}
\end{figure}

The dust color temperature of sources in the field of IRS 1 was
obtained from the ratio of images taken at 7.9 and 18.2
$\micron$. Both images were registered by matching the position of
source \#50 and zoomed to show the sources, the resulting image is
shown in the Figure \ref{ctemp}. Sources \#50 and \#48 are clearly
seen and they have color temperatures of $\sim280$ ($\pm$10) K and
$\sim215$ ($\pm$15) K respectively. Sources \#60 and \#60b are not
detected in this map due to their low S/N ratio.

The dust color temperature of sources \#95 and IRS 1:SE were also
obtained, but in this case through the fluxes measured in the 12.5 and
18.2 $\micron$ band, since the image taken at 7.9 $\micron$ has low
S/N ratio. The results are: T = 270 ($\pm$10) K for source \#95, T =
100 ($\pm$20) K for source IRS 1:SE. The uncertainties in the temperatures
represent the difference between the results obtained using the 7.9/18.2
$\micron$ flux ratio and the 12.5/18.2 $\micron$ flux ratio.

MIR luminosity of sources \#48 and \#50 were estimated by integrating the fluxes
measured at 3.5 ($L$ band), 7.9, 12.5 and 18.2 $\micron$. The MIR luminosity of
source \#60b, instead, was estimated by performing the same integration, but
using only the fluxes measured at 7.9, 12.5 and 18.2 $\micron$. From these
results, we can estimate the bolometric luminosity assuming that MIR luminosity
must represent $\sim10$\% of the bolometric luminosity \citep{church89a}. The
results are presented in the Table \ref{spt}. The spectral types were obtained
from the derived bolometric luminosities and the grids of stellar models of
\citet{scha92}, and they should be considered approximate given the large
bolometric correction used to derive the bolometric luminosity. 

The silicate absorption map can be used to investigate the spatial
distribution of dust and was obtained by ratioing the 9.8
$\micron$ band image and a ``continuum" image. This ``continuum" was
produced by averaging the images at 7.9 and 12.5 $\micron$. The
silicate optical depth, obtained by the procedure described by
\citet{gezari98}, is $\tau _{9.8}$ = 3.7 for source \#50,
corresponding to $A_{V}=59$ mag, assuming $A_{V}=16\times\tau_{9.8}$
\citep{rieke85} and $\tau_{9.8}$ = 0.77 for source \#48 corresponding
to an $A_{V}$ = 12. Previous estimates of the opacity of IRS 1 are
from \citet{persson76} who found $\tau_{9.8}$=3.5 using a 15$\arcsec$
beamwidth and from \citet{persi94}, who found $\tau_{9.8}\simeq4.7$
using a CVF 10 $\micron$ spectrum taken with an aperture of
7.5$\arcsec$. Both estimates were made using large apertures which
include contributions from all sources and also their surrounding
dust. Values for visual extinction obtained with NIR data for source
\#48 are $A_{V}=14.3$ (FBDC) and $A_{V}=11$ \citep{persi87}.

\subsection{Where is the ionizing source of NGC 3576?}

IRS 1 was initially thought to be an important source of radiation to
NGC 3576, but we have found that it does not contribute substantially,
as we show below. The number of Lyman continuum photons derived from
the radio data \citep{depree99} is $\sim10^{50}$ s$^{-1}$ for NGC
3576. A single O3 star or a cluster of at least 10 O6 stars is needed
to produce the amount of Lyman continuum photons found in NGC 3576,
but where are the ionizing sources?

\begin{figure}
\epsscale{1}
\plotone{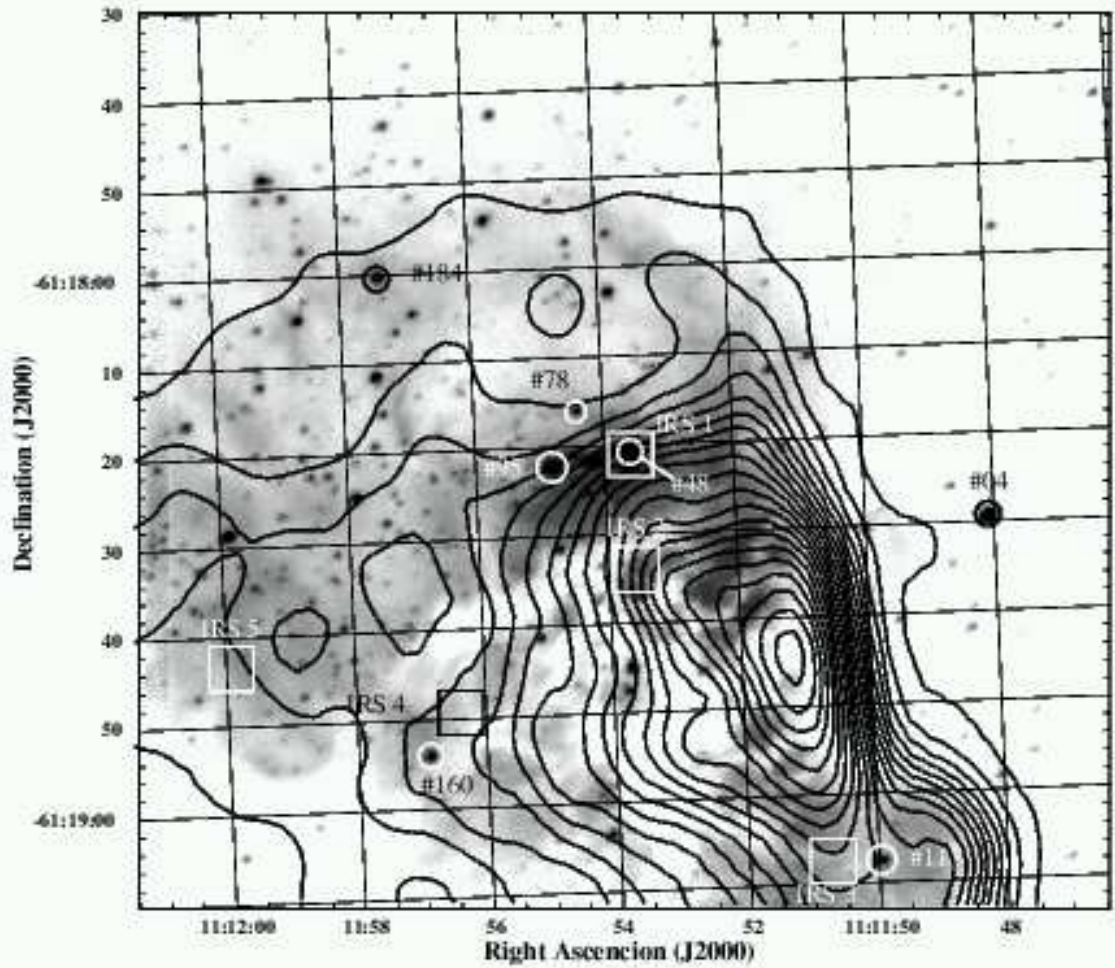}
\caption{Composite image of NGC 3576. The $K$-band image from FBDC is in
inverted gray scale. The 3.4 cm continuum contour curves are from the
radio image of NGC~3576 (beamwidth of $7\arcsec\times7\arcsec$), kindly
provided by C. DePree. Circles indicate the YSO candidates which show
featureless spectra from FBDC and boxes are the MIR sources from
\citet{persson76}. Contour levels are between 30 and 4100 mJy beam$^{-1}$.
\label{3576-cmp}}
\end{figure}

Figure \ref{3576-cmp} can give us an idea. The $K$-band image of NGC
3576, from FBDC, was overplotted by the 3.4 cm continuum contour
curves, from \citet{depree99}. The radio beamwidth in the image is
$7\arcsec\times7\arcsec$. YSOs are shown inside the circles and the
positions of MIR sources, detected by \citet{persson76}, are shown as
boxes. None of the MIR sources (IRS 1-5) is associated with the radio
peak emission, moreover IRS 1 does not affect the level curves in its
vicinity. However, the strong continuum emission and rather
large beam size of the radio image may be limiting our ability to detect
compact radio emission around the IRS 1 source. The radio data neither
eliminate nor preclude their presence. However, according to our torus model
and expectations for the evolutionary status of the massive YSOs outlined
in \S1, we expect these sources might be in the UCHII region
phase. Despite the lack of a compact 3.4 cm source, there is very
strong Br$\gamma$ emission surrounding the IRS 1 sources \citep{blum03}. This
strongly suggests a local contribution to the ionization of the
circumstellar environment of these objects. For the brightest $K-$band
source, \# 48, we have recently obtained a high spectral resolution
($R=$50,000) spectrum at 2.17 $\mu$m on the Gemini South telescope
which shows a marked double peak morphology. This morphology does not
include broad wings typical of a disk like signature, but would be
consistent with a shell or torus geometry. This spectrum (to be
published in an upcoming paper) is consistent with our torus model,
and we thus prefer the interpretation that the IRS 1 sources are in
an UCHII region phase, rather than in an earlier phase of evolution.

Returning to the issue of where the dominant ionizing sources in
NGC~3576 lie, we note that the radio peak emission corresponds to a
region of dark patches to the south in the $K-$band image and for this
reason no ionizing source(s) could be detected at $K$. The radio peak emission
lies far from any MIR source and it is identified as an UC\ion{H}{2} region in
the figure 2(f) of \citet{walsh01}. The ionizing
source(s) must be a stellar cluster, just blocked by the dark clouds
in the line of sight. In this case the cluster stars must be somewhat
evolved: they already have broken out their birth material producing
the quoted Lyman continuum photons that are ripping apart the
interviewing dark patches and hence do not form any detectable
UC\ion{H}{2} region. If the stars in the cluster were in an earlier, more
enshrouded stage of evolution they could not produce the observed radio
emission. A complete MIR map of the region is planned in order to further
study the massive YSO candidates and their environment on the same basis as
we have done for IRS 1.

\section{Summary and Conclusions}

We have presented MIR images of NGC 3576. IRS 1 is resolved into 4
sources for the first time at 10 $\micron$. The brightest source in
the $N$ band is source \#50, which is different from the situation seen
at $K$, where source \#48 is the brightest. We have also presented MIR
images of the YSO candidate \#95, and the detection of NIR sources
\#52, \#54 and \#73 as a negative object in the sky beam.

SEDs of sources \#48, \#50 and \#60 were constructed from 1.25 to 18.2
$\micron$ by combining the data available in the literature with our
data. We also constructed the MIR SEDs for sources \#48, \#50, \#95
and the detected companion of source \#60, named \#60b, based on the
fluxes measured through narrow-band filters.

The optical depth, and hence the visual extinction toward each object
was obtained from the silicate absorption feature and amounts to
$\tau_{9.8}$ = 3.7 (A$_{V}$ = 59 mag) for \#50 and $\tau_{9.8}$ = 0.77
(A$_{V}$ = 12 mag) for \#48. Previous values of visual extinction for
source \#48 are $A_{V} = 14$ (FBDC) and $A_{V} = 11$
\citep{persi87}.

The MIR luminosity of sources \#48, \#50, \#60b and \#95 were obtained
after integrating their MIR fluxes in order to give an approximate of their
bolometric luminosities and hence the spectral types. We
argue, based on the radio continuum image, that the IRS 1 sources
are not the major contributors to the ionization of NGC 3576. Nevertheless,
we conclude that sources \#48, \#50 and \#95 are young massive stars
possibly surrounded by a torus of gas and dust which is responsible
for the different SEDs observed. 

We report the detection of a new diffuse MIR source without
any NIR counterpart.  This source was found at 8$\arcsec$\ southeast
of \#50 and has a core-halo morphology. We named it IRS 1:SE and
derived the color temperature of $\sim$100 K. IRS 1:SE has both morphology and
color temperature compatible with HCs \citep{osorio99}. Moreover its emission
profile is indistinguishable from a point source embedded in extended emission.
For these reasons we do not believe IRS 1:SE is just a bright knot in the tail
of the extended emission of source \#50. However, additional data are
needed to give any firm conclusion.

The ionizing source of NGC 3576 is believed to be behind the dark clouds seen
in the $K$-band image, blocked from detection at shorter wavelengths by large
amounts of intervening dust.

\acknowledgments

The authors would like to thank Cris De Pree for kindly provided the
radio image of NGC 3576 and an anonymous referee whose comments improved the
clarity of the presentation.  C. L. B. an A. D. thank FAPESP (under the
grant 00/03462-5), PRONEX and PROAP for financial support. P. S. C. appreciates
continuous support from the National Science Foundation.

\clearpage

\clearpage

\clearpage

\clearpage

\clearpage

\clearpage

\clearpage

\clearpage

\clearpage

\clearpage

\begin{deluxetable}{ccccccc}
\tablewidth{0pt}
\tablecaption{\label{stdlog}Filter characteristics and summary of the 
observations}
\tablehead{
\colhead{} & \colhead{Central} &
\colhead{} &
\colhead{OSCIR} &
\colhead{OSCIR} &
\colhead{Observed} &
\colhead{Exposure}\\
\colhead{Filter} & \colhead{Wavelength} &
\colhead{Bandwidth} &
\colhead{ZMFD\tablenotemark{a}} &
\colhead{FD $\alpha$ CMa\tablenotemark{b}} &
\colhead{PSF} &
\colhead{Time\tablenotemark{c}}\\
\colhead{} & \colhead{(\micron)} &
\colhead{(\micron)} &
\colhead{(Jy)} &
\colhead{(Jy)} &
\colhead{(arcsec)} &
\colhead{(s)}
}
\startdata
7.9 & 7.91 & 0.755 & 59.4 & 207.01 & 0.4 &43\\
9.8 & 9.80 & 0.952 & 39.9 & 138.03 & 0.4 & 43\\
N & 10.75 & 5.23 & 37.8 & 131.82 & 0.5 & 40\\
12.5 & 12.49 & 1.156 & 25.1 & 87.09 & 0.4 & 43\\
IHW18 & 18.17 & 1.651 & 11.9 & 40.88 & 0.65 & 40\\
\enddata

\tablenotetext{a}{Zero Magnitude Flux Density}
\tablenotetext{b}{Flux Density of Sirius through OSCIR filters, assuming
$N$=-1.35 mag. and $Q$=-1.34 mag. \citet{cohen92}.}
\tablenotetext{c}{On source integration time.}

\end{deluxetable}

\clearpage

\begin{deluxetable}{lccccccccc}
\tabletypesize{\footnotesize}
\tablewidth{0pt}
\tablecaption{NIR and MIR fluxes of IRS 1 in the broad--band filters. The 
(\nodata) entries indicate non detections
in the filter. $JHK$ fluxes are from FBDC, L flux from \citet{moneti92} and $N$ flux from this work,
see text.\label{nir-mir}}
\tablehead{
\colhead{Source} &
\multicolumn{2}{c}{J} &
\multicolumn{2}{c}{H} &
\multicolumn{2}{c}{K} &
\multicolumn{2}{c}{L} &
\colhead{N} \\
\colhead{} &
\colhead{(mag)} & \colhead{(mJy)} &
\colhead{(mag)} & \colhead{(mJy)} &
\colhead{(mag)} & \colhead{(mJy)} &
\colhead{(mag)} & \colhead{(mJy)} &
\colhead{(mJy)}
}
\startdata
\#48 & 13.6 & 5.86 & 10.56 & 62.50 & 8.35 & 299 & 5.52 & 1790 & 4950 ($\pm 495$)\\
\#50 & \nodata & \nodata & \nodata & \nodata & 11.07 & 24.40 & 4.05 & 6920 & 33540 ($\pm 3350$)\\
\#60 & 12.96 & 10.60 & 11.89 & 18.4 & 11.06 & 24.70 & \nodata & \nodata & 913 ($\pm 100$)\\
\#60b & \nodata & \nodata & \nodata & \nodata & \nodata & \nodata & \nodata & \nodata & 1346 ($\pm 135$)\\
\enddata

\end{deluxetable}

\clearpage

\begin{deluxetable}{ccccc}
\tablewidth{0pt}
\tablecaption{\label{shifts}Relative positions of sources detected in the field
of IRS 1 in the $K$ and $N$ bands. The coordinates are in arcsec relative to
source \#50, $\alpha$= 11h11m53.62s $\delta$=
-61$\arcdeg$18$\arcmin$21.9$\arcsec$ (J2000)}
\tablehead{
\colhead{} &
\multicolumn{2}{c}{K} &
\multicolumn{2}{c}{N} \\
\colhead{Source} &
\colhead{$\Delta$RA} & \colhead{$\Delta$Dec} &
\colhead{$\Delta$RA} & \colhead{$\Delta$Dec} \\
\colhead{} &
\colhead{($\pm0.1\arcsec$)} & \colhead{($\pm0.1\arcsec$)} &
\colhead{($\pm0.1\arcsec$)} & \colhead{($\pm0.1\arcsec$)}
}
\startdata
\#48 & 1.0 & 1.5 & 1.0 & 1.5\\
\#60 & -1.1 & 1.8 & -1.2 & 1.8\\
\#60b & -0.5 & 1.5 & -0.7 & 1.6\\
\enddata

\end{deluxetable}

\clearpage

\begin{deluxetable}{ccccc}
\tablewidth{0pt}
\tablecaption{\label{mir}MIR fluxes of detected sources in narrow--band filters. (\nodata) indicate
that no image was taken in the band.}
\tablehead{
\colhead{Source} & 
\colhead{7.9 (mJy)} &
\colhead{9.8 (mJy)} &
\colhead{12.5 (mJy)} &
\colhead{IHW18 (mJy)}
}
\startdata
\#48 & 3360 ($\pm340$) & 1880 ($\pm190$) & 4520 ($\pm450$) & 9340 ($\pm930$)\\
\#50 & 36560 ($\pm3650$) & 1410 ($\pm140$)& 51600 ($\pm5160$) & 35560 ($\pm3550$)\\
\#52 & \nodata & \nodata & 227 ($\pm23$)& \nodata\\
\#54 & \nodata & \nodata & $<$130 & \nodata\\
\#60 & $<$430 & 764 ($\pm 77$) & $<$200 & $<$8730\tablenotemark{b} \\
\#60b & $<$430 & 907 ($\pm91$) & 1510 ($\pm150$)& $<$8730\tablenotemark{b} \\
\#73 & \nodata & \nodata & 173 ($\pm20$) & \nodata\\
\#95 & 151 ($\pm38$) & 198 ($\pm20$)& 184 ($\pm20$)& $<$890\\
IRS 1:SE\tablenotemark{a} & $<$160 & 322 ($\pm33$)& 498 ($\pm50$)& 2460 ($\pm245$)\\
\enddata
\tablenotetext{a}{Values for IRS 1:SE are in mJy arcsec$^{-2}$.}
\tablenotetext{b}{8730 mJy is the combined 18.2 $\micron$ flux from both source,
\#60 and \#60b, as these were not resolved at this wavelength.}

\end{deluxetable}

\clearpage

\begin{deluxetable}{ccccccc}
\tablewidth{0pt}
\tablecaption{\label{spt}Color temperature of dust and spectral types derived from the integrated 
MIR fluxes for the YSO candidates.}
\tablehead{
\colhead{Source} &
\colhead{T$_{c}$} &
\colhead{L$_{MIR}$} &
\colhead{$\log{L/L_{\sun}}$\tablenotemark{a}} &
\colhead{Sp. Type} &
\colhead{Spectral} \\
\colhead{} &
\colhead{(K)} &
\colhead{(erg s$^{-1}$)} &
\colhead{} &
\colhead{ZAMS} &
\colhead{Index}
}
\startdata
\#48 & 215 ($\pm10$)& 2.7$\times10^{36}$ & 3.8 & B1 & 0.6 \\
\#50 & 280 ($\pm10$)& 2.4$\times10^{37}$ & 4.8 & O8 & 3.4\\
\#60b & $>$100\tablenotemark{b} & 3.5$\times10^{35}$ & 2.9 & B3 & \nodata\\
\#95 & 270 ($\pm15$)& 1.5$\times10^{35}\tablenotemark{c}$ & 2.5 & B5 & -0.8 \\
\enddata
\tablenotetext{a}{Bolometric luminosity estimated by adopting a ``bolometric
correction" as given by \citet{church89a} for UCHII regions. L = 10 $\times$ L$_{MIR}$,
see text.}
\tablenotetext{b}{Lower limit for the temperature due to the unresolved 
nature of source at 18.2 $\micron$}
\tablenotetext{c}{MIR luminosity is an upper limit based on the upper limit to the flux
in the 18 $\micron$ band. See Table 3.}

\end{deluxetable}

\end{document}